

\expandafter\ifx\csname phyzzx\endcsname\relax
 \message{It is better to use PHYZZX format than to
          \string\input\space PHYZZX}\else
 \wlog{PHYZZX macros are already loaded and are not
          \string\input\space again}%
 \endinput \fi
\catcode`\@=11 
\let\rel@x=\relax
\let\n@expand=\relax
\def\pr@tect{\let\n@expand=\noexpand}
\let\protect=\pr@tect
\let\gl@bal=\global
%
%
%
\newfam\cpfam
\newdimen\b@gheight             \b@gheight=12pt
\newcount\f@ntkey               \f@ntkey=0
\def\f@m{\afterassignment\samef@nt\f@ntkey=}
\def\samef@nt{\fam=\f@ntkey \the\textfont\f@ntkey\rel@x}
\def\setstr@t{\setbox\strutbox=\hbox{\vrule height 0.85\b@gheight
                                depth 0.35\b@gheight width\z@ }}
\input phyzzx.fonts
%
\def\rm{\n@expand\f@m0 }
\def\mit{\n@expand\f@m1 }         
\def\cal{\n@expand\f@m2 }
\def\it{\n@expand\f@m\itfam}
\def\sl{\n@expand\f@m\slfam}
\def\bf{\n@expand\f@m\bffam}
\def\tt{\n@expand\f@m\ttfam}
\def\caps{\n@expand\f@m\cpfam}    
\def\em@{\rel@x\ifnum\f@ntkey=0 \it \else
        \ifnum\f@ntkey=\bffam \it \else \rm \fi \fi }
\def\em{\n@expand\em@}
\def\fourteenpoint{\fourteenf@nts \samef@nt \b@gheight=14pt \setstr@t }
\def\twelvepoint{\twelvef@nts \samef@nt \b@gheight=12pt \setstr@t }
\def\tenpoint{\tenf@nts \samef@nt \b@gheight=10pt \setstr@t }
\normalbaselineskip = 19.2pt plus 0.2pt minus 0.1pt 
\normallineskip = 1.5pt plus 0.1pt minus 0.1pt
\normallineskiplimit = 1.5pt
\newskip\normaldisplayskip
\normaldisplayskip = 14.4pt plus 3.6pt minus 10.0pt 
\newskip\normaldispshortskip
\normaldispshortskip = 6pt plus 5pt
\newskip\normalparskip
\normalparskip = 6pt plus 2pt minus 1pt
\newskip\skipregister
\skipregister = 5pt plus 2pt minus 1.5pt
\newif\ifsingl@
\newif\ifdoubl@
\newif\iftwelv@  \twelv@true
\def\singlespace{\singl@true\doubl@false\spaces@t}
\def\doublespace{\singl@false\doubl@true\spaces@t}
\def\normalspace{\singl@false\doubl@false\spaces@t}
\def\Tenpoint{\tenpoint\twelv@false\spaces@t}
\def\Twelvepoint{\twelvepoint\twelv@true\spaces@t}
\def\spaces@t{\rel@x
      \iftwelv@ \ifsingl@\subspaces@t3:4;\else\subspaces@t1:1;\fi
       \else \ifsingl@\subspaces@t3:5;\else\subspaces@t4:5;\fi \fi
      \ifdoubl@ \multiply\baselineskip by 5
         \divide\baselineskip by 4 \fi }
\def\subspaces@t#1:#2;{
      \baselineskip = \normalbaselineskip
      \multiply\baselineskip by #1 \divide\baselineskip by #2
      \lineskip = \normallineskip
      \multiply\lineskip by #1 \divide\lineskip by #2
      \lineskiplimit = \normallineskiplimit
      \multiply\lineskiplimit by #1 \divide\lineskiplimit by #2
      \parskip = \normalparskip
      \multiply\parskip by #1 \divide\parskip by #2
      \abovedisplayskip = \normaldisplayskip
      \multiply\abovedisplayskip by #1 \divide\abovedisplayskip by #2
      \belowdisplayskip = \abovedisplayskip
      \abovedisplayshortskip = \normaldispshortskip
      \multiply\abovedisplayshortskip by #1
        \divide\abovedisplayshortskip by #2
      \belowdisplayshortskip = \abovedisplayshortskip
      \advance\belowdisplayshortskip by \belowdisplayskip
      \divide\belowdisplayshortskip by 2
      \smallskipamount = \skipregister
      \multiply\smallskipamount by #1 \divide\smallskipamount by #2
      \medskipamount = \smallskipamount \multiply\medskipamount by 2
      \bigskipamount = \smallskipamount \multiply\bigskipamount by 4 }
\def\normalbaselines{ \baselineskip=\normalbaselineskip
   \lineskip=\normallineskip \lineskiplimit=\normallineskip
   \iftwelv@\else \multiply\baselineskip by 4 \divide\baselineskip by 5
     \multiply\lineskiplimit by 4 \divide\lineskiplimit by 5
     \multiply\lineskip by 4 \divide\lineskip by 5 \fi }
\Twelvepoint  
\interlinepenalty=50
\interfootnotelinepenalty=5000
\predisplaypenalty=9000
\postdisplaypenalty=500
\hfuzz=1pt
\vfuzz=0.2pt
\newdimen\HOFFSET  \HOFFSET=0pt
\newdimen\VOFFSET  \VOFFSET=0pt
\newdimen\HSWING   \HSWING=0pt
\dimen\footins=8in
%
%
%
\newskip\pagebottomfiller
\pagebottomfiller=\z@ plus \z@ minus \z@
\def\pagecontents{
   \ifvoid\topins\else\unvbox\topins\vskip\skip\topins\fi
   \dimen@ = \dp255 \unvbox255
   \vskip\pagebottomfiller
   \ifvoid\footins\else\vskip\skip\footins\footrule\unvbox\footins\fi
   \ifr@ggedbottom \kern-\dimen@ \vfil \fi }
\def\makeheadline{\vbox to 0pt{ \skip@=\topskip
      \advance\skip@ by -12pt \advance\skip@ by -2\normalbaselineskip
      \vskip\skip@ \line{\vbox to 12pt{}\the\headline} \vss
      }\nointerlineskip}
\def\makefootline{\baselineskip = 1.5\normalbaselineskip
                 \line{\the\footline}}
\newif\iffrontpage
\newif\ifp@genum
\def\nopagenumbers{\p@genumfalse}
\def\pagenumbers{\p@genumtrue}
\pagenumbers
\newtoks\paperheadline
\newtoks\paperfootline
\newtoks\letterheadline
\newtoks\letterfootline
\newtoks\letterinfo
\newtoks\date
\paperheadline={\hfil}
\paperfootline={\hss\iffrontpage\else\ifp@genum\tenrm\folio\hss\fi\fi}
\letterheadline{\iffrontpage \hfil \else
    \rm \ifp@genum page~~\folio\fi \hfil\the\date \fi}
\letterfootline={\iffrontpage\the\letterinfo\else\hfil\fi}
\letterinfo={\hfil}
\def\monthname{\rel@x\ifcase\month 0/\or January\or February\or
   March\or April\or May\or June\or July\or August\or September\or
   October\or November\or December\else\number\month/\fi}
\def\today{\monthname~\number\day, \number\year}
\date={\today}
\headline=\paperheadline 
\footline=\paperfootline 
\countdef\pageno=1      \countdef\pagen@=0
\countdef\pagenumber=1  \pagenumber=1
\def\advancepageno{\gl@bal\advance\pagen@ by 1
   \ifnum\pagenumber<0 \gl@bal\advance\pagenumber by -1
    \else\gl@bal\advance\pagenumber by 1 \fi
    \gl@bal\frontpagefalse  \swing@ }
\def\folio{\ifnum\pagenumber<0 \romannumeral-\pagenumber
           \else \number\pagenumber \fi }
\def\swing@{\ifodd\pagenumber \gl@bal\advance\hoffset by -\HSWING
             \else \gl@bal\advance\hoffset by \HSWING \fi }
\def\footrule{\dimen@=\prevdepth\nointerlineskip
   \vbox to 0pt{\vskip -0.25\baselineskip \hrule width 0.35\hsize \vss}
   \prevdepth=\dimen@ }
\let\footnotespecial=\rel@x
\newdimen\footindent
\footindent=24pt
\def\Textindent#1{\noindent\llap{#1\enspace}\ignorespaces}
\def\Vfootnote#1{\insert\footins\bgroup
   \interlinepenalty=\interfootnotelinepenalty \floatingpenalty=20000
   \singl@true\doubl@false\Tenpoint
   \splittopskip=\ht\strutbox \boxmaxdepth=\dp\strutbox
   \leftskip=\footindent \rightskip=\z@skip
   \parindent=0.5\footindent \parfillskip=0pt plus 1fil
   \spaceskip=\z@skip \xspaceskip=\z@skip \footnotespecial
   \Textindent{#1}\footstrut\futurelet\next\fo@t}

\def\vfootnote#1{\Vfootnote{${#1}$}}
\def\footnote#1{\attach{#1}\vfootnote{#1}}

\def\foot{\attach\footsymbolgen\vfootnote{\footsymbol}}
\let\footsymbol=\star
\newcount\lastf@@t           \lastf@@t=-1
\newcount\footsymbolcount    \footsymbolcount=0
\newif\ifPhysRev
\def\footsymbolgen{\bumpfootsymbolcount \generatefootsymbol \footsymbol }
\def\bumpfootsymbolcount{\rel@x
   \iffrontpage \bumpfootsymbolpos \else \advance\lastf@@t by 1
     \ifPhysRev \bumpfootsymbolneg \else \bumpfootsymbolpos \fi \fi
   \gl@bal\lastf@@t=\pagen@ }
\def\bumpfootsymbolpos{\ifnum\footsymbolcount <0
                            \gl@bal\footsymbolcount =0 \fi
    \ifnum\lastf@@t<\pagen@ \gl@bal\footsymbolcount=0
     \else \gl@bal\advance\footsymbolcount by 1 \fi }
\def\bumpfootsymbolneg{\ifnum\footsymbolcount >0
             \gl@bal\footsymbolcount =0 \fi
         \gl@bal\advance\footsymbolcount by -1 }
\def\fd@f#1 {\xdef\footsymbol{\mathchar"#1 }}
\def\generatefootsymbol{\ifcase\footsymbolcount \fd@f 13F \or \fd@f 279
        \or \fd@f 27A \or \fd@f 278 \or \fd@f 27B \else
        \ifnum\footsymbolcount <0 \fd@f{023 \number-\footsymbolcount }
         \else \fd@f 203 {\loop \ifnum\footsymbolcount >5
                \fd@f{203 \footsymbol } \advance\footsymbolcount by -1
                \repeat }\fi \fi }

\def\nonfrenchspacing{\sfcode`\.=3001 \sfcode`\!=3000 \sfcode`\?=3000
        \sfcode`\:=2000 \sfcode`\;=1500 \sfcode`\,=1251 }
\nonfrenchspacing
\newdimen\d@twidth
{\setbox0=\hbox{s.} \gl@bal\d@twidth=\wd0 \setbox0=\hbox{s}
        \gl@bal\advance\d@twidth by -\wd0 }
\def\removehglue{\loop \unskip \ifdim\lastskip >\z@ \repeat }
\def\roll@ver#1{\removehglue \nobreak \count255 =\spacefactor \dimen@=\z@
        \ifnum\count255 =3001 \dimen@=\d@twidth \fi
        \ifnum\count255 =1251 \dimen@=\d@twidth \fi
    \iftwelv@ \kern-\dimen@ \else \kern-0.83\dimen@ \fi
   #1\spacefactor=\count255 }
\def\step@ver#1{\rel@x \ifmmode #1\else \ifhmode
        \roll@ver{${}#1$}\else {\setbox0=\hbox{${}#1$}}\fi\fi }
\def\attach#1{\step@ver{\strut^{\mkern 2mu #1} }}
%
%
%
\newcount\chapternumber      \chapternumber=0
\newcount\sectionnumber      \sectionnumber=0
\newcount\equanumber         \equanumber=0
\let\chapterlabel=\rel@x
\let\sectionlabel=\rel@x
\newtoks\chapterstyle        \chapterstyle={\Number}
\newtoks\sectionstyle        \sectionstyle={\chapterlabel.\Number}
\newskip\chapterskip         \chapterskip=\bigskipamount
\newskip\sectionskip         \sectionskip=\medskipamount
\newskip\headskip            \headskip=8pt plus 3pt minus 3pt
\newdimen\chapterminspace    \chapterminspace=15pc
\newdimen\sectionminspace    \sectionminspace=10pc
\newdimen\referenceminspace  \referenceminspace=20pc
\def\chapterreset{\gl@bal\advance\chapternumber by 1
   \ifnum\equanumber<0 \else\gl@bal\equanumber=0\fi
   \sectionnumber=0 \let\sectionlabel=\rel@x
   {\pr@tect\xdef\chapterlabel{\the\chapterstyle{\the\chapternumber}}}}
\def\alphabetic#1{\count255='140 \advance\count255 by #1\char\count255}
\def\Alphabetic#1{\count255='100 \advance\count255 by #1\char\count255}
\def\Roman#1{\uppercase\expandafter{\romannumeral #1}}
\def\roman#1{\romannumeral #1}
\def\Number#1{\number #1}
\def\BLANC#1{}
\def\titleparagraphs{\interlinepenalty=9999
     \leftskip=0.03\hsize plus 0.22\hsize minus 0.03\hsize
     \rightskip=\leftskip \parfillskip=0pt
     \hyphenpenalty=9000 \exhyphenpenalty=9000
     \tolerance=9999 \pretolerance=9000
     \spaceskip=0.333em \xspaceskip=0.5em }
\def\titlestyle#1{\par\begingroup \titleparagraphs
     \iftwelv@\fourteenpoint\else\twelvepoint\fi
   \noindent #1\par\endgroup }
\def\spacecheck#1{\dimen@=\pagegoal\advance\dimen@ by -\pagetotal
   \ifdim\dimen@<#1 \ifdim\dimen@>0pt \vfil\break \fi\fi}
\def\chapter#1{\par \penalty-300 \vskip\chapterskip
   \spacecheck\chapterminspace
   \chapterreset \titlestyle{\S\chapterlabel.~#1}
   \nobreak\vskip\headskip \penalty 30000
   {\pr@tect\wlog{\string\chapter\space \chapterlabel}} }

\def\section#1{\par \ifnum\the\lastpenalty=30000\else
   \penalty-200\vskip\sectionskip \spacecheck\sectionminspace\fi
   \gl@bal\advance\sectionnumber by 1
   {\pr@tect
   \xdef\sectionlabel{\the\sectionstyle\the\sectionnumber}
   \wlog{\string\section\space \sectionlabel}}
   \noindent {\caps\enspace\sectionlabel.~~#1}\par
   \nobreak\vskip\headskip \penalty 30000 }
\def\subsection#1{\par
   \ifnum\the\lastpenalty=30000\else \penalty-100\smallskip \fi
   \noindent\undertext{#1}\enspace \vadjust{\penalty5000}}

\def\undertext#1{\vtop{\hbox{#1}\kern 1pt \hrule}}
\def\ACK{\par\penalty-100\medskip \spacecheck\sectionminspace
   \line{\fourteenrm\hfil Acknowledgements\hfil}\nobreak\vskip\headskip }

\def\APPENDIX#1#2{\par\penalty-300\vskip\chapterskip
   \spacecheck\chapterminspace \chapterreset \xdef\chapterlabel{#1}
   \titlestyle{Appendix #2} \nobreak\vskip\headskip \penalty 30000
   \wlog{\string\Appendix~\chapterlabel} }
\def\Appendix#1{\APPENDIX{#1}{#1}}
\def\appendix{\APPENDIX{A}{}}
\def\unnumberedchapters{\let\makechapterlabel=\rel@x
      \let\chapterlabel=\rel@x  \sectionstyle={\BLANC}
      \let\sectionlabel=\rel@x \sequentialequations }
%
%
%
\def\eqname#1{\rel@x {\pr@tect
  \ifnum\equanumber<0 \xdef#1{{\rm(\number-\equanumber)}}%
     \gl@bal\advance\equanumber by -1
  \else \gl@bal\advance\equanumber by 1
     \ifx\chapterlabel\rel@x \def\d@t{}\else \def\d@t{.}\fi
    \xdef#1{{\rm(\chapterlabel\d@t\number\equanumber)}}\fi #1}}
\def\eq{\eqname\?}
\def\eqn{\eqno\eqname}

\def\eqinsert#1{\noalign{\dimen@=\prevdepth \nointerlineskip
   \setbox0=\hbox to\displaywidth{\hfil #1}
   \vbox to 0pt{\kern 0.5\baselineskip\hbox{$\!\box0\!$}\vss}
   \prevdepth=\dimen@}}
%

%
%
\def\GENITEM#1;#2{\par \hangafter=0 \hangindent=#1
    \Textindent{$ #2 $}\ignorespaces}
\outer\def\newitem#1=#2;{\gdef#1{\GENITEM #2;}}

\newdimen\itemsize                \itemsize=30pt
\newitem\item=1\itemsize;
\newitem\sitem=1.75\itemsize;     
\newitem\ssitem=2.5\itemsize;     
\outer\def\newlist#1=#2&#3&#4;{\toks0={#2}\toks1={#3}%
   \count255=\escapechar \escapechar=-1
   \alloc@0\list\countdef\insc@unt\listcount     \listcount=0
   \edef#1{\par
      \countdef\listcount=\the\allocationnumber
      \advance\listcount by 1
      \hangafter=0 \hangindent=#4
      \Textindent{\the\toks0{\listcount}\the\toks1}}
   \expandafter\expandafter\expandafter
    \edef\c@t#1{begin}{\par
      \countdef\listcount=\the\allocationnumber \listcount=1
      \hangafter=0 \hangindent=#4
      \Textindent{\the\toks0{\listcount}\the\toks1}}
   \expandafter\expandafter\expandafter
    \edef\c@t#1{con}{\par \hangafter=0 \hangindent=#4 \noindent}
   \escapechar=\count255}
\def\c@t#1#2{\csname\string#1#2\endcsname}
\newlist\point=\Number&.&1.0\itemsize;
\newlist\subpoint=(\alphabetic&)&1.75\itemsize;
\newlist\subsubpoint=(\roman&)&2.5\itemsize;
%

%
%
%
%
\newcount\referencecount     \referencecount=0
\newcount\lastrefsbegincount \lastrefsbegincount=0
\newif\ifreferenceopen       \newwrite\referencewrite
\newdimen\refindent          \refindent=30pt
\def\normalrefmark#1{\attach{\scriptscriptstyle [ #1 ] }}
\let\PRrefmark=\attach
\def\NPrefmark#1{\step@ver{{\;[#1]}}}
\def\refmark#1{\rel@x\ifPhysRev\PRrefmark{#1}\else\normalrefmark{#1}\fi}
\def\refend@{\refmark{\number\referencecount}}
\def\refend{\refend@{}\space }
\def\refsend{\refmark{\count255=\referencecount
   \advance\count255 by-\lastrefsbegincount
   \ifcase\count255 \number\referencecount
   \or \number\lastrefsbegincount,\number\referencecount
   \else \number\lastrefsbegincount-\number\referencecount \fi}\space }
\def\REFNUM#1{\rel@x \gl@bal\advance\referencecount by 1
    \xdef#1{\the\referencecount }}
\def\Refnum#1{\REFNUM #1\refend@ } 
\def\REF#1{\REFNUM #1\R@FWRITE\ignorespaces}
\def\Ref#1{\Refnum #1\REFWRITE }
\def\ref{\Ref\?}
\def\REFS#1{\REFNUM #1\gl@bal\lastrefsbegincount=\referencecount
    \REFWRITE }

\def\r@fitem#1{\par \hangafter=0 \hangindent=\refindent \Textindent{#1}}
\def\refitem#1{\r@fitem{#1.}}
\def\NPrefitem#1{\r@fitem{[#1]}}
\def\NPrefs{\let\refmark=\NPrefmark \let\refitem=\NPrefitem}
\def\REFWRITE{\R@FWRITE\rel@x }
\def\R@FWRITE#1{\ifreferenceopen \else \gl@bal\referenceopentrue
     \immediate\openout\referencewrite=\jobname.refs
     \toks@={\begingroup \refoutspecials \catcode`\^^M=10 }%
     \immediate\write\referencewrite{\the\toks@}\fi
    \immediate\write\referencewrite{\noexpand\refitem %
                                    {\the\referencecount}}%
    \p@rse@ndwrite \referencewrite #1}
\begingroup
 \catcode`\^^M=\active \let^^M=\relax %
 \gdef\p@rse@ndwrite#1#2{\begingroup \catcode`\^^M=12 \newlinechar=`\^^M%
         \chardef\rw@write=#1\sc@nlines#2}%
 \gdef\sc@nlines#1#2{\sc@n@line \g@rbage #2^^M\endsc@n \endgroup #1}%
 \gdef\sc@n@line#1^^M{\expandafter\toks@\expandafter{\deg@rbage #1}%
         \immediate\write\rw@write{\the\toks@}%
         \futurelet\n@xt \sc@ntest }%
\endgroup
\def\sc@ntest{\ifx\n@xt\endsc@n \let\n@xt=\rel@x
       \else \let\n@xt=\sc@n@notherline \fi \n@xt }
\def\sc@n@notherline{\sc@n@line \g@rbage }
\def\deg@rbage#1{}
\let\g@rbage=\relax    \let\endsc@n=\relax
\def\refout{\par\penalty-400\vskip\chapterskip
   \spacecheck\referenceminspace
   \ifreferenceopen \Closeout\referencewrite \referenceopenfalse \fi
   \line{\fourteenrm\hfil References\hfil}\vskip\headskip
   \input \jobname.refs
   }
\def\refoutspecials{\sfcode`\.=1000 \interlinepenalty=1000
         \rightskip=\z@ plus 1em minus \z@ }
\def\Closeout#1{\toks0={\par\endgroup}\immediate\write#1{\the\toks0}%
   \immediate\closeout#1}
%
%
\newcount\figurecount     \figurecount=0
\newcount\tablecount      \tablecount=0
\newif\iffigureopen       \newwrite\figurewrite
\newif\iftableopen        \newwrite\tablewrite
\def\FIGNUM#1{\rel@x \gl@bal\advance\figurecount by 1
    \xdef#1{\the\figurecount}}
\def\FIGURE#1{\FIGNUM #1\F@GWRITE\ignorespaces }
\let\FIG=\FIGURE

\def\figitem#1{\r@fitem{#1)}}
\def\FIGWRITE{\F@GWRITE\rel@x }
\def\TABNUM#1{\rel@x \gl@bal\advance\tablecount by 1
    \xdef#1{\the\tablecount}}
\def\TABLE#1{\TABNUM #1\T@BWRITE\ignorespaces }

\def\tabitem#1{\r@fitem{#1:}}
\def\TABWRITE{\T@BWRITE\rel@x }
\def\F@GWRITE#1{\iffigureopen \else \gl@bal\figureopentrue
     \immediate\openout\figurewrite=\jobname.figs
     \toks@={\begingroup \catcode`\^^M=10 }%
     \immediate\write\figurewrite{\the\toks@}\fi
    \immediate\write\figurewrite{\noexpand\figitem %
                                 {\the\figurecount}}%
    \p@rse@ndwrite \figurewrite #1}
\def\T@BWRITE#1{\iftableopen \else \gl@bal\tableopentrue
     \immediate\openout\tablewrite=\jobname.tabs
     \toks@={\begingroup \catcode`\^^M=10 }%
     \immediate\write\tablewrite{\the\toks@}\fi
    \immediate\write\tablewrite{\noexpand\tabitem %
                                 {\the\tablecount}}%
    \p@rse@ndwrite \tablewrite #1}
\def\figout{\par\penalty-400
   \vskip\chapterskip\spacecheck\referenceminspace
   \iffigureopen \Closeout\figurewrite \figureopenfalse \fi
   \line{\fourteenrm\hfil FIGURE CAPTIONS\hfil}\vskip\headskip
   \input \jobname.figs
   }
\def\tabout{\par\penalty-400
   \vskip\chapterskip\spacecheck\referenceminspace
   \iftableopen \Closeout\tablewrite \tableopenfalse \fi
   \line{\fourteenrm\hfil TABLE CAPTIONS\hfil}\vskip\headskip
   \input \jobname.tabs
   }
%
%
%
\newbox\picturebox
\def\p@cht{\ht\picturebox }
\def\p@cwd{\wd\picturebox }
\def\p@cdp{\dp\picturebox }
\newdimen\xshift
\newdimen\yshift
\newdimen\captionwidth
\newskip\captionskip
\captionskip=15pt plus 5pt minus 3pt
\def\fullwidth{\captionwidth=\hsize }
\newtoks\Caption
\newif\ifcaptioned
\newif\ifselfcaptioned
\def\caption{\captionedtrue \Caption }
\newcount\linesabove
\newif\iffileexists
\newtoks\picfilename
\def\fil@#1 {\fileexiststrue \picfilename={#1}}
\def\file#1{\if=#1\let\n@xt=\fil@ \else \def\n@xt{\fil@ #1}\fi \n@xt }
\def\pl@t{\begingroup \pr@tect
    \setbox\picturebox=\hbox{}\fileexistsfalse
    \let\height=\p@cht \let\width=\p@cwd \let\depth=\p@cdp
    \xshift=\z@ \yshift=\z@ \captionwidth=\z@
    \Caption={}\captionedfalse
    \linesabove =0 \picturedefault }
\def\plot{\pl@t \selfcaptionedfalse }
\def\Picture#1{\gl@bal\advance\figurecount by 1
    \xdef#1{\the\figurecount}\pl@t \selfcaptionedtrue }

\def\s@vepicture{\iffileexists \parsefilename \redopicturebox \fi
   \ifdim\captionwidth>\z@ \else \captionwidth=\p@cwd \fi
   \xdef\lastpicture{\iffileexists
        \setbox0=\hbox{\raise\the\yshift \vbox{%
              \moveright\the\xshift\hbox{\picturedefinition}}}%
        \else \setbox0=\hbox{}\fi
         \ht0=\the\p@cht \wd0=\the\p@cwd \dp0=\the\p@cdp
         \vbox{\hsize=\the\captionwidth \line{\hss\box0 \hss }%
              \ifcaptioned \vskip\the\captionskip \noexpand\Tenpoint
                \ifselfcaptioned Figure~\the\figurecount.\enspace \fi
                \the\Caption \fi }}%
    \endgroup }
\let\endpicture=\s@vepicture
\def\savepicture#1{\s@vepicture \global\let#1=\lastpicture }
\def\displaypicture{\fullwidth \s@vepicture $$\lastpicture $${}}
\def\toppicture{\fullwidth \s@vepicture \topinsert
    \lastpicture \medskip \endinsert }
\def\midpicture{\fullwidth \s@vepicture \midinsert
    \lastpicture \endinsert }
%
%
\def\leftpicture{\pres@tpicture
    \dimen@i=\hsize \advance\dimen@i by -\dimen@ii
    \setbox\picturebox=\hbox to \hsize {\box0 \hss }%
    \wr@paround }
\def\rightpicture{\pres@tpicture
    \dimen@i=\z@
    \setbox\picturebox=\hbox to \hsize {\hss \box0 }%
    \wr@paround }
\def\pres@tpicture{\gl@bal\linesabove=\linesabove
    \s@vepicture \setbox\picturebox=\vbox{
         \kern \linesabove\baselineskip \kern 0.3\baselineskip
         \lastpicture \kern 0.3\baselineskip }%
    \dimen@=\p@cht \dimen@i=\dimen@
    \advance\dimen@i by \pagetotal
    \par \ifdim\dimen@i>\pagegoal \vfil\break \fi
    \dimen@ii=\hsize
    \advance\dimen@ii by -\parindent \advance\dimen@ii by -\p@cwd
    \setbox0=\vbox to\z@{\kern-\baselineskip \unvbox\picturebox \vss }}
\def\wr@paround{\Caption={}\count255=1
    \loop \ifnum \linesabove >0
         \advance\linesabove by -1 \advance\count255 by 1
         \advance\dimen@ by -\baselineskip
         \expandafter\Caption \expandafter{\the\Caption \z@ \hsize }%
      \repeat
    \loop \ifdim \dimen@ >\z@
         \advance\count255 by 1 \advance\dimen@ by -\baselineskip
         \expandafter\Caption \expandafter{%
             \the\Caption \dimen@i \dimen@ii }%
      \repeat
    \edef\n@xt{\parshape=\the\count255 \the\Caption \z@ \hsize }%
    \par\noindent \n@xt \strut \vadjust{\box\picturebox }}
\let\picturedefault=\relax
\let\parsefilename=\relax
\def\redopicturebox{\let\picturedefinition=\rel@x
   \errhelp=\disabledpictures
   \errmessage{This version of TeX cannot handle pictures.  Sorry.}}
\newhelp\disabledpictures
     {You will get a blank box in place of your picture.}
%
%
%
%
%
%
%
%
%
%
\def\FRONTPAGE{\ifvoid255\else\vfill\penalty-20000\fi
   \gl@bal\pagenumber=1     \gl@bal\chapternumber=0
   \gl@bal\equanumber=0     \gl@bal\sectionnumber=0
   \gl@bal\referencecount=0 \gl@bal\figurecount=0
   \gl@bal\tablecount=0     \gl@bal\frontpagetrue
   \gl@bal\lastf@@t=0       \gl@bal\footsymbolcount=0}

\def\papers{\papersize\headline=\paperheadline\footline=\paperfootline}
\def\papersize{
   \advance\hoffset by\HOFFSET \advance\voffset by\VOFFSET
   \pagebottomfiller=0pc
   \skip\footins=\bigskipamount \normalspace }
\papers  
%
%
\newskip\lettertopskip       \lettertopskip=20pt plus 50pt
\newskip\letterbottomskip    \letterbottomskip=\z@ plus 100pt
\newskip\signatureskip       \signatureskip=40pt plus 3pt
\def\lettersize{\hsize=6.5in \vsize=8.5in \hoffset=0in \voffset=0.5in
   \advance\hoffset by\HOFFSET \advance\voffset by\VOFFSET
   \pagebottomfiller=\letterbottomskip
   \skip\footins=\smallskipamount \multiply\skip\footins by 3
   \singlespace }
\def\MEMO{\lettersize \headline=\letterheadline \footline={\hfil }%
   \let\rule=\memorule \FRONTPAGE \memohead }

\def\memodate{\afterassignment\MEMO \date }
\def\memit@m#1{\smallskip \hangafter=0 \hangindent=1in
    \Textindent{\caps #1}}
\def\subject{\memit@m{Subject:}}
\def\topic{\memit@m{Topic:}}
\def\from{\memit@m{From:}}
\def\memorule{\medskip\hrule height 1pt\bigskip}  
\def\memohead{\centerline{\fourteenrm MEMORANDUM}}
\newwrite\labelswrite
\newtoks\rw@toks
\def\letters{\lettersize
   \headline=\letterheadline \footline=\letterfootline
   \immediate\openout\labelswrite=\jobname.lab}

\let\letterhead=\rel@x
\def\addressee#1{\medskip\line{\hskip 0.75\hsize plus\z@ minus 0.25\hsize
                               \the\date \hfil }%
   \vskip \lettertopskip
   \ialign to\hsize{\strut ##\hfil\tabskip 0pt plus \hsize \crcr #1\crcr}
   \writelabel{#1}\medskip \noindent\hskip -\spaceskip \ignorespaces }
\def\rwl@begin#1\cr{\rw@toks={#1\crcr}\rel@x
   \immediate\write\labelswrite{\the\rw@toks}\futurelet\n@xt\rwl@next}
\def\rwl@next{\ifx\n@xt\rwl@end \let\n@xt=\rel@x
      \else \let\n@xt=\rwl@begin \fi \n@xt}
\let\rwl@end=\rel@x
\def\writelabel#1{\immediate\write\labelswrite{\noexpand\labelbegin}
     \rwl@begin #1\cr\rwl@end
     \immediate\write\labelswrite{\noexpand\labelend}}
\newtoks\FromAddress         \FromAddress={}
\newtoks\sendername          \sendername={}
\newbox\FromLabelBox
\newdimen\labelwidth          \labelwidth=6in
\def\makelabels{\afterassignment\Makelabels \sendersname=}
\def\Makelabels{\FRONTPAGE \letterinfo={\hfil } \MakeFromBox
     \immediate\closeout\labelswrite  \input \jobname.lab\vfil\eject}
\let\labelend=\rel@x
\def\labelbegin#1\labelend{\setbox0=\vbox{\ialign{##\hfil\cr #1\crcr}}
     \MakeALabel }
\def\MakeFromBox{\gl@bal\setbox\FromLabelBox=\vbox{\Tenpoint
     \ialign{##\hfil\cr \the\sendername \the\FromAddress \crcr }}}
\def\MakeALabel{\vskip 1pt \hbox{\vrule \vbox{
        \hsize=\labelwidth \hrule\bigskip
        \leftline{\hskip 1\parindent \copy\FromLabelBox}\bigskip
        \centerline{\hfil \box0 } \bigskip \hrule
        }\vrule } \vskip 1pt plus 1fil }
\def\signed#1{\par \nobreak \bigskip \dt@pfalse \begingroup
  \everycr={\noalign{\nobreak
            \ifdt@p\vskip\signatureskip\gl@bal\dt@pfalse\fi }}%
  \tabskip=0.5\hsize plus \z@ minus 0.5\hsize
  \halign to\hsize {\strut ##\hfil\tabskip=\z@ plus 1fil minus \z@\crcr
          \noalign{\gl@bal\dt@ptrue}#1\crcr }%
  \endgroup \bigskip }
\newbox\letterb@x
\def\lettertext{\par \vskip\parskip \unvcopy\letterb@x \par }
\def\multiletter{\setbox\letterb@x=\vbox\bgroup
      \everypar{\vrule height 1\baselineskip depth 0pt width 0pt }
      \singlespace \topskip=\baselineskip }
\def\letterend{\par\egroup}
%
%
%
\newskip\frontpageskip
\newtoks\Pubnum   
\newtoks\Pubtype  \let\pubtype=\Pubtype
\newif\ifp@bblock  \p@bblocktrue
\def\PH@SR@V{\doubl@true \baselineskip=24.1pt plus 0.2pt minus 0.1pt
             \parskip= 3pt plus 2pt minus 1pt }
\def\PHYSREV{\papers\PhysRevtrue\PH@SR@V}

\def\titlepage{\FRONTPAGE\papers\ifPhysRev\PH@SR@V\fi
   \ifp@bblock\p@bblock \else\hrule height\z@ \rel@x \fi }
\def\nopubblock{\p@bblockfalse}
\def\endpage{\vfil\break}
\frontpageskip=12pt plus .5fil minus 2pt
\Pubtype={}
\Pubnum={}
\def\p@bblock{\begingroup \tabskip=\hsize minus \hsize
   \baselineskip=1.5\ht\strutbox \topspace-2\baselineskip
   \halign to\hsize{\strut ##\hfil\tabskip=0pt\crcr
       \the\Pubnum\crcr\the\date\crcr\the\pubtype\crcr}\endgroup}
\def\title#1{\vskip\frontpageskip \titlestyle{#1} \vskip\headskip }
\def\author#1{\vskip\frontpageskip\titlestyle{\twelvecp #1}\nobreak}

\def\address#1{\par\kern 5pt\titlestyle{\twelvepoint\it #1}}
\def\andaddress{\par\kern 5pt \centerline{\sl and} \address}

\def\abstract{\par\dimen@=\prevdepth \hrule height\z@ \prevdepth=\dimen@
   \vskip\frontpageskip\centerline{\fourteenrm Abstract}\vskip\headskip }

%
%
%

\def\\{\rel@x \ifmmode \backslash \else {\tt\char`\\}\fi }
\def\sequentialequations{\rel@x \if\equanumber<0 \else
  \gl@bal\equanumber=-\equanumber \gl@bal\advance\equanumber by -1 \fi }
\def\nextline{\unskip\nobreak\hfill\break}

\def\journal#1&#2(#3){\begingroup \let\journal=\dummyj@urnal
    \unskip, \sl #1\unskip~\bf\ignorespaces #2\rm
    (\afterassignment\j@ur \count255=#3), \endgroup\ignorespaces }
\def\j@ur{\ifnum\count255<100 \advance\count255 by 1900 \fi
          \number\count255 }
\def\dummyj@urnal{%
    \toks@={Reference foul up: nested \journal macros}%
    \errhelp={Your forgot & or ( ) after the last \journal}%
    \errmessage{\the\toks@ }}

\def\topspace{\hrule height 0pt depth 0pt \vskip}

\def\partder#1#2{{\partial #1\over\partial #2}}
\def\Buildrel#1\under#2{\mathrel{\mathop{#2}\limits_{#1}}}
\def\becomes#1{\mathchoice{\becomes@\scriptstyle{#1}}
   {\becomes@\scriptstyle{#1}} {\becomes@\scriptscriptstyle{#1}}
   {\becomes@\scriptscriptstyle{#1}}}
\def\becomes@#1#2{\mathrel{\setbox0=\hbox{$\m@th #1{\,#2\,}$}%
        \mathop{\hbox to \wd0 {\rightarrowfill}}\limits_{#2}}}
\def\bra#1{\left\langle #1\right|}
\def\ket#1{\left| #1\right\rangle}

\def\VEV#1{\left\langle #1\right\rangle}

\def\Tr{\mathop{\rm Tr}\nolimits}

\let\int=\intop         \let\oint=\ointop
\def\lsim{\mathrel{\mathpalette\@versim<}}
\def\gsim{\mathrel{\mathpalette\@versim>}}
\def\@versim#1#2{\vcenter{\offinterlineskip
        \ialign{$\m@th#1\hfil##\hfil$\crcr#2\crcr\sim\crcr } }}
\def\big#1{{\hbox{$\left#1\vbox to 0.85\b@gheight{}\right.\n@space$}}}
\def\Big#1{{\hbox{$\left#1\vbox to 1.15\b@gheight{}\right.\n@space$}}}
\def\bigg#1{{\hbox{$\left#1\vbox to 1.45\b@gheight{}\right.\n@space$}}}
\def\Bigg#1{{\hbox{$\left#1\vbox to 1.75\b@gheight{}\right.\n@space$}}}
\def\){\mskip 2mu\nobreak }
%
%
%
\let\sec@nt=\sec
\def\sec{\rel@x\ifmmode\let\n@xt=\sec@nt\else\let\n@xt\section\fi\n@xt}
\def\obsolete#1{\message{Macro \string #1 is obsolete.}}
\def\firstsec#1{\obsolete\firstsec \section{#1}}
\def\firstsubsec#1{\obsolete\firstsubsec \subsection{#1}}
\def\thispage#1{\obsolete\thispage \gl@bal\pagenumber=#1\frontpagefalse}
\def\thischapter#1{\obsolete\thischapter \gl@bal\chapternumber=#1}
\def\splitout{\obsolete\splitout\rel@x}
\def\prop{\obsolete\prop \propto }
\def\nextequation#1{\obsolete\nextequation \gl@bal\equanumber=#1
   \ifnum\the\equanumber>0 \gl@bal\advance\equanumber by 1 \fi}
\def\BOXITEM{\afterassigment\B@XITEM\setbox0=}
\def\B@XITEM{\par\hangindent\wd0 \noindent\box0 }
%
%
%
\def\phyzzx{PHY\setbox0=\hbox{Z}\copy0 \kern-0.5\wd0 \box0 X}
        
\everyjob{\xdef\today{\monthname~\number\day, \number\year}
        \input myphyx.tex }
\message{ by V.K.}
%
\catcode`\@=12 
%


\def\bal{$\alpha$\kern-.7em\hbox{$\alpha$}}

\sequentialequations
\Pubnum{EPHOU 94-001}
\titlepage
\title{\centerline{Charged and Neutral Vortex Excitations in a Mean
        Field Theory}
       \nextline\centerline{for the Fractional Quantum Hall Effect}}
\author{N{\tenrm obuki} M{\tenrm AEDA}
\foot{maeda@particle.phys.hokudai.ac.jp}
}

\address{Department of Physics, Hokkaido University, Sapporo 060
JAPAN
}
\abstract
Applying a bi-local mean field approximation to the fractional
quantum Hall state of $\nu=1/3$, we obtain charged and neutral vortex
mean field solutions numerically. We calculate the mean field energy
and the fluctuation corrections. The charged vortex has a fractional
charge and a fractional angular momentum.
The neutral vortex is a bound state of two charged vortices and has
a zero charge and a zero angular momentum. The creation energy of the
neutral vortex is about a half of the pair creation energy of two
charged vortices. The magnetic field dependence
of the gap energy agrees with the Laughlin's quasiparticle gap energy.

\endpage

\doublespace


\REF\bc{
{\it The Quantum Hall Effect} ed. R.E.Prange and S.M.Girvin
(Springer, New York, 1990).}
\REF\Lau{R.B.Laughlin,Phys.Rev.Lett.{\bf50}(1983),1395.}
\REF\a{K.Ishikawa,Prog.Theor.Phys,
Suppl.No.107(1992),167;Vol.88(1992),881. }
\REF\A{K.Ishikawa and N.Maeda,Phys.Lett.A{\bf186}(1994),65.}
\REF\B{K.Ishikawa and N.Maeda,Prog.Theor.Phys.{\bf91}(1994),237.}
\REF\C{A.K.Kerman,S.Levit and T.Troudet,Ann.Phys.(N.Y.){\bf 148}
(1983),436.}
\REF\abo{A.Abrikosov,ZhETF(USSR){\bf32}(1957),1442
[JETP(Sov.Phys.){\bf5}(1957),1174].\nextline
H.Nielsen and P.Olesen,Nucl.Phys.B61(1973),45.}
\REF\rj{R.Morf and B.I.Halperin,
Phys.Rev.B33(1986),2221.}
\REF\rc{S.M.Girvin,A.H.MacDonald and P.M.Platzman,
Phys.Rev.Lett.{\bf 54}(1985),581;
\nextline Phys.Rev.B33(1986),2481.}
\REF\rb{D.Yoshioka,B.I.Halperin and P.A.Lee,Phys.Rev.Lett.
{\bf 50}(1983),1219.}
\REF\ra{F.D.M.Haldane and E.H.Rezayi,Phys.Rev.Lett.
{\bf 54}(1985),237.}
\REF\rg{R.L.Willet,H.L.Stormer,D.C.Tsui,A.C.Gossard
and J.H.English,\nextline Phys.Rev.B37(1988),8476.}
\REF\pdpw{A.Pinczek,B.S.Dennis,L.N.Pfeiffer and K.West,
Phys.Rev.Lett.{\bf70}(1993),3983.}

\chapter{Introduction}
We consider the two-dimensional interacting spinless electron
system in a perpendicular magnetic field.
The fractional quantum Hall effect(FQHE)
$^{\bc)}$ can be characterized
by the existence of an energy gap in a partially filled Landau
level. Strong correlations among electrons with the Coulomb
interaction are responsible for the energy gap.
Since a partially filled many-electron state in the Landau level
has an enormous degeneracy, systematic calculations based on
a microscopic theory are very difficult.
Laughlin proposed a phenomenological many-body wave function for
a ground state of an incompressible quantum liquid$^{\Lau)}$.
In the Laughlin's theory the excited state includes quasiholes
and quasiparticles which have fractional charges and obey the
fractional statistics. The pair creation energy of a quasihole
and a quasiparticle gives a finite energy gap and is in a good
agreement with experimental values at strong magnetic fields.
Although the Laughlin's theory explains the FQHE
phenomenologically, it may be desirable to have more systematic
calculational method.

Recently a bi-local mean field theory
$^{\a),\A)}$ was applied to the
many-body system for the FQHE and obtained a uniform mean
field solution for a ground state$^{\a)}$
and charged vortex solutions
for a topological excited state$^{\B)}$.
It was found that
the vortex solutions have a fractional
charge and a fractional angular momentum and
have vorticity one.
In this theory the correlation function
$\langle \psi^\dagger(y)\psi(x) \rangle$ is treated as a mean
field. It was found that the gap energy is rather smaller
in a mean field theory than the
Laughlin's one.
In the present paper we calculate the fluctuation corrections to the
vortex energy and find that the gap energy becomes close to the
Laughlin's one. We also calculate charged and neutral vortex
solutions of vorticity 2. The neutral vortex solution has a zero
charge and zero angular momentum.
The neutral vortex is considered to be a bound state of two charged
vortices which have charges of opposite sign.
The neutral vortex energy is about a half of the pair creation energy
of two charged vortices.

In \S 2, we review a bi-local mean field theory for the
FQHE. We point out that we need to change the self-consistent
condition for a systematic expansion around the vortex
mean field solutions slightly. Nevertheless, it is shown in appendix
that our results
obtained in this paper is not changed.
In \S 3, we consider the ansatz forms of charged vortex
solutions of vorticity $n_{\rm v}$ and get the exact relation between
the fractional charge and fractional angular momentum of vortex in
the large $B$ limit.
We also consider a neutral vortex of vorticity 2. The fluctuation
corrections to the vortex energy are calculated in \S 4.
The results of numerical calculations are presented in \S 5.
In \S 6, we summarize our results.

\chapter{Mean field approximation for the FQHE}
We consider the two-dimensional electron system in a perpendicular
uniform magnetic field $B$ which is described by a Hamiltonian
$$
H(\psi^\dagger,\psi)
=\int d^2x[\psi^\dagger(x){({\bf P}+e{\bf A})^2\over2m}\psi(x)
+{1\over2}\int d^2y\psi^\dagger(x)\psi^\dagger(y)V(x-y)
\psi(y)\psi(x)],
\eqno\eq
$$
where
$
P_i=-i\partder{}{x^i},\
{\bf A}={B\over2}(-x_2,x_1)
$ and
$
V(r)=e^2/(\kappa r).\
$
We use the unit $\hbar=c=1$ and fix $\kappa=13$, $m=0.07m_e$ for
GaAs.
We supposed that the magnetic field $B$ is strong enough to make
electrons fully polarized and neglected the spin effect of electrons.

In the functional integral formalism,
the partition function is written as
$$
Z=\Tr(e^{-\beta H(\hat\psi^\dagger,\hat\psi)})
=\int {\cal D}\psi^\dagger{\cal D}\psi \exp[-\int\nolimits
^\beta\nolimits_0 dt(\int d^2x
\psi^\dagger
\partial_t\psi+H(\psi^\dagger,\psi))],
\eqno\eq
$$
where $\psi^\dagger$ and $\psi$ are anti-commuting field variables,
$\hat\psi^\dagger$ and $\hat\psi$ are anti-commuting field operators
and $\beta=1/k_B T$.
We introduce a bi-local
auxiliary field $U(x,y;t)$ and convert the interaction term into
a quadratic form of $\psi^\dagger$ and $\psi$ as
$$
\eqalign{
H_U(\psi^\dagger,\psi)&=\int d^2x\psi^\dagger(x;t)
{({\bf P}+e{\bf A})^2\over2m}\psi(x;t)\cr
&+{1\over2}\int d^2xd^2y V(x-y)[U(x,y;t)U(y,x;t)\cr
&\qquad -U(x,y;t)\psi^\dagger(x;t)\psi(y;t)-
U(y,x;t)\psi^\dagger(y;t)\psi(x;t)],
}
\eqno\eq
$$
$$
Z={\cal N}^{-1}\int{\cal D}\psi^\dagger{\cal D}\psi{\cal D}U
\exp[-\int\nolimits^\beta\nolimits_0 dt(\int d^2x\psi^\dagger
\partial_t\psi+H_U(\psi^\dagger,\psi))]
$$
$$
={\cal N}^{-1}\int {\cal D}\psi^\dagger{\cal D}\psi
{\cal D}U \exp[-S(\psi^\dagger,\psi,U)]
={\cal N}^{-1}\int {\cal D}U \exp[-S_{\rm eff}(U)],
\eqno\eq
$$
$$
{\cal N}=\int {\cal D}U\exp[-\int\nolimits
^\beta\nolimits_0 dt(\int d^2x d^2y {V(x-y)
\over 2}U(x,y;t)U(y,x;t))].
\eqno\eq
$$
In the mean field approximation, the partition function is calculated
at the stationary point of $S_{\rm eff}$ as
$$
Z_0=e^{-S_{\rm eff}(U_0)}
=\Tr(\exp[-\beta H_{U_0}(
\hat\psi^\dagger,\hat\psi)]),
\eqno\eq
$$
$$
\partder{S_{\rm eff}}{U}\biggr\vert_{U=U_0}=0.
\eqno\eq
$$
Using a correlation function, Eq.(2.7) can be written as
$$
U_0(x,y)=\langle\psi^\dagger(y)\psi(x)\rangle_{U_0},
\eqno\eq
$$
$$
\VEV{\psi^\dagger(y)\psi(x)}_{U_0}\equiv Z^{-1}_0
\int{\cal D}\psi^\dagger
{\cal D}\psi(\psi^\dagger(y)\psi(x))\exp[-S(\psi^\dagger,\psi,U_0)]
$$
$$
=Z^{-1}_0\Tr(\hat\psi^\dagger(y)\hat\psi(x)\exp[-\beta H_{U_0}(
\hat\psi^\dagger,\hat\psi)])
\rightarrow
\langle E_0\vert \hat\psi^\dagger(y)\hat\psi(x)\vert E_0\rangle,
\eqno\eq
$$
in the limit $\beta\rightarrow\infty$.
The state
$\vert E_0\rangle$ is the lowest energy eigenstate of $H_{U_0}(
\hat\psi^\dagger,\hat\psi)$. We solve the self-consistency condition
Eq.(2.8) under the following ansatz form :
$$
U_0(x,y)=U_0\rho(x,y)e^{-\gamma(x-y)^2}\exp[i\int\nolimits
^y\nolimits_x \alpha_i(\xi)d\xi^i],
\eqno\eq
$$
where $U_0={\nu/\pi R^2_0},\ \gamma={1/2R^2_0},\ R_0=\sqrt{2/eB}$
and $\rho(x,y)=\rho(y,x)$. The line integral
is along a straight line between $x$ and $y$. $\rho(x,y)$ and $
\alpha_i(\xi)$ are unknown real functions and we will determine
these functions under appropriate assumptions.
We assume that the mean field $U_0(x,y)$ coincides with a uniform
self-consistent solution at infinity :
$$
\left\{\matrix{
\lim\limits_{x\rightarrow\infty}\rho(x,x)=1,\cr
\lim\limits_{\xi\rightarrow\infty}\alpha_i(\xi)=eA_i(\xi).\cr}
\right.
\eqno\eq
$$
Equations
(2.10) and (2.11) mean that the electron density $U_0(x,x)$ becomes
uniform density of the filling factor $\nu$ at infinity.
By substituting Eq.(2.10) into Eq.(2.3), we get a mean field Hamiltonian :
$$
\eqalign{
H_{U_0}&=\int d^2x
\psi^\dagger(x)h_0\psi(x)+{U_0^2\over 2}\int d^2xd^2y V(x-y)
\rho^2(x,y)e^{-2\gamma(x-y)^2},\cr
h_0&=[{({\bf P}+e{\bf A})^2\over2m}-
F(({\bf P}+\hbox{\bal})^2)\rho({\bar x},x)]_{\bar x\rightarrow x},\cr
F(p^2)&=e^2U_0{\pi^{3/2}\over\kappa\gamma^{1/2}}e^{-p^2/8\gamma}
I_0(p^2/8\gamma), \cr
}
\eqno\eq
$$
where $I_0$ is a zero-th order modified Bessel function.
$h_0$ is a single-particle Hamiltonian and $[\ \ ]
_{\bar x\rightarrow x}$
represents to take a limit $\bar x\rightarrow x$ after ${\bf P}$
operates on a function of $x$.

It is convenient to expand the electron field operator by the $h_0$'s
eigenfunctions. We assume that the system has a rotational invariance
and the eigenfunctions are classified by the angular momentum quantum
number, $l$, as

$$
\eqalign{
\hat\psi(x)&=\sum_{l,n}u_{l,n}(x){\hat a}_{l,n},\
\int d^2xu^*_{l,n}(x)u_{l',n'}(x)=\delta_{l,l'}\delta_{n,n'},\cr
 h_0 u_{l,n}(x)&=E_{l,n}u_{l,n}(x),
\ \{{\hat a}^\dagger_{l,n},{\hat a}_{l',n'}\}
=\delta_{l,l'}\delta_{n,n'},\cr
E_{l,0}<&E_{l,1}<E_{l,2}<\cdots,
}
\eqno\eq
$$
where ${\hat a}^\dagger_{l,n}$ and ${\hat a}_{l',n'}$ are anti-commuting
creation and annihilation operators.
The N-electron state is defined by a totally anti-symmetric tensor
$F_{l_1,\cdots l_N}$ as
$$
\eqalign{
\ket{\Psi}&=\sum_{l_1,l_2\cdots l_N=0}^M F_{l_1,\cdots l_N}
{\hat a}^{\dagger}_{l_1,0}{\hat a}^{\dagger}_{l_2,0}\cdots
{\hat a}^{\dagger}_{l_N,0}\ket{0},\cr
{\hat a}_{l,n}\ket{0}&=0\quad{\rm for\ any\ }l,\ n,}
\eqno\eq
$$
where $M$ is a maximum value of the angular momentum of a
single-particle. Since we concentrate
upon the case of $\nu<1$ (especially
$\nu=1/3$), we restrict $\ket{\Psi}$ within the $n=0$ eigenstates in
Eq.(2.14). For a free electron system the $n=0$ eigenstates belong to
the lowest Landau level. $F_{l_1,\cdots l_N}$ is zero unless
$l_1+l_2+\cdots+l_N=L_T$ for a fixed total angular momentum $L_T$.
{}From this property one can prove the following relations :
$$
\eqalign{
(N\cdot N!)&
\sum_{l_2\cdots l_N=0}^M F^*_{n,l_2,\cdots l_N}F_{m,l_2,\cdots l_N}
=\nu_n\delta_{n,m},\quad 0\leq \nu_n \leq 1,\cr
&\sum_{l=0}^M \nu_l =N.}
\eqno\eq
$$
We don't need to know the explicit form of $F_{l_1,\cdots l_N}$
but need to know the value of $\nu_l$ for the following calculations.
By Eqs.(2.14) and (2.15) we can calculate the correlation function as
$$
\bra{\Psi}\hat\psi^\dagger(y)\hat\psi(x)\ket{\Psi}
=\sum_{l=0}^M \nu_l u^*_{l,0}(y)u_{l,0}(x).
\eqn\ps
$$
The condition that the electron density becomes a uniform density
of filling factor $\nu$ means $\lim\limits
_{l\rightarrow\infty}\nu_l=\nu=N/(M+1)$
for $N\rightarrow\infty$.
Using Eqs.(2.10) and (2.16), the self-consistency condition (2.8)
becomes
$$
U_0\rho(x,y)e^{-\gamma(x-y)^2}\exp[i\int\nolimits
^y\nolimits_x \alpha_i(\xi)d\xi^i]=\sum_{l=0}^M
\nu_l u^*_{l,0}(y)u_{l,0}(x).
\eqno\eq
$$

Generally the N-electron state $\ket{\Psi}$ in Eq.(2.14) is a
superposition of different ${}_{M+1}C_N$ states.
Hence the different states in $\ket{\Psi}$ have to be degenerate
in the energy eigenvalue because the lowest energy eigenstate
survives in the self-consistency condition (2.8) as Eq.(2.9).
We will show later that the energy eigenvalue $E_{l,0}$ is
degenerate for a large $l$, but is not degenerate for a small $l$
in the vortex mean field solution.
Therefore we need to modify the self-consistency condition (2.8)
so as to make the eigenvalues degenerate. In fact this is possible
by using an arbitrariness in the limiting procedure of discretized time
slices$^{\C)}$.
The details are summarized in appendix. Since the results
obtained by using Eq.(2.8) become the same results as
obtained by using a modified
one, we continue to use Eq.(2.8) in the following.

\chapter{Vortex mean field solutions}
Following Ref.5), we work on the ansatz form of a vortex as
$$
\eqalign{
\alpha_i(x)&=eA_i(x)+n_i(x),\quad
{\bf n}(x)={n(r)\over r^2}(-x_2,x_1),\cr
}
\eqno\eq
$$
with the boundary conditions :
$$
\eqalign{
&n(0)=n_{\rm v}=1,2,3,\cdots,\cr
&\lim_{r\rightarrow\infty}n(r)=0,\cr
&\rho(x,0)=0,\ \rho(x,x)=\rho(r).}
\eqno\eq
$$
In Eqs.(3.1) and (3.2), the form of $n_i(x)$ is a reminiscent of the
vortex solution in the Abelian Higgs theory
$^{\abo)}$.
It is seen by separating the non-singular part ${\tilde n}_i$ in
$n_i$ as
$$
\eqalign{
\int\nolimits_x\nolimits
^y &n_i(\xi)d\xi^i=n_{\rm v}(\theta_y-\theta_x)+
\int\nolimits_x\nolimits^y {\tilde n}_i(\xi)d\xi^i,\cr
&{\tilde n}_i(x)={n(r)-n(0)\over r^2}(-x_2,x_1),\cr
\oint_{r=\infty}&{\tilde n}_i(\xi)d\xi^i=n_{\rm v}.
}
\eqno\eq
$$
The singular part of $n_i$ gives a phase dependence $e^{in_{\rm v}
\theta}$
to the expectation value of the field operator and the non-singular
part gives a dynamical magnetic field, $\vert\nabla\times\tilde{\bf n}
\vert={1\over r}{dn\over dr}$,
localized around a vortex.
$n_{\rm v}$ is called vorticity and characterize the
topological excitations.

The uniform density mean field solution was already obtained in Ref.3)
as
$$
\eqalign{
&\rho(x,y)=1,\cr
&\alpha_i(x)=eA_i(x),\cr
&\nu_l=\nu\quad{\rm for\ }l=0,1,\cdots.}
\eqno\eq
$$
We denote the uniform density state as
$$
\eqalign{&
\ket{\Psi_0}=\sum_{l_1,l_2\cdots l_N=0}^M F^{(0)}_{l_1,\cdots l_N}
{\hat a}^{\dagger}_{l_1,0}{\hat a}^{\dagger}_{l_2,0}\cdots
{\hat a}^{\dagger}_{l_N,0}\ket{0},\cr
&(N\cdot N!)
\sum_{l_2\cdots l_N=0}^M F^{(0)*}_{n,l_2,\cdots l_N}
F^{(0)}_{m,l_2,\cdots l_N}
=\nu\delta_{n,m}.}
\eqno\eq
$$
The explicit form of $F^{(0)}_{l_1,\cdots l_N}$ is obtained from the
coefficient $F^{P}_{l_1,\cdots l_N}$ of the Laughlin wave function
as $\prod_{i<j}(z_i-z_j)^P=\sum_{l_1\cdots l_N}F^{P}_{l_1,\cdots l_N}
z^{l_1}\cdots z^{l_N}$ for odd integer $P=1/\nu$.
The eigenfunction of $h_0$ for the uniform density mean field is
written as
$$
\eqalign{
u_{l,0}^{(0)}&(x)=(\pi R_0^2 l!)^{-1/2}{z^*}^l e^{
-{z^*}z/2},\ l=0,1,2,\cdots,\cr
&z=(x_1+ix_2)/R_0,}
\eqno\eq
$$
and the energy eigenvalue is degenerate as
$$
E_{l,0}^{(0)}={eB\over 2m}-F(eB).
\eqno\eq
$$

\noindent
(A) {\it Charged vortex}

Using the uniform density state we construct a quasihole state of a
charge $-n_{\rm v}\nu e$ as
$$
\ket{\Psi_{-n_{\rm v}\nu}}
=\sum_{l_1,l_2\cdots l_N=0}^M F^{(0)}_{l_1,\cdots l_N}
{\hat a}^{\dagger}_{l_1+n_{\rm v},0}{\hat a}^{\dagger}_
{l_2+n_{\rm v},0}\cdots
{\hat a}^{\dagger}_{l_N+n_{\rm v},0}\ket{0}.
\eqno\eq
$$
Note that ${\hat a}^{\dagger}$ in Eq.(3.5)
and ${\hat a}^{\dagger}$
in Eq.(3.8) are not same because ${\hat a}^{\dagger}$ is dependent upon
the
set of eigenfunctions of $h_0$, $\{u_{l,n}(x)\}$, and $h_0$ is
dependent upon
the state of electrons.
The state $\ket{\Psi_{-n_{\rm v}\nu}}$ gives $\nu_0=\cdots=\nu_{
n_{\rm v}-1}=0$ and $\nu_{n_{\rm v}}=\nu_{n_{\rm v}+1}=\cdots=\nu$.
In fact the charge difference from the uniform density state is
$$
\bra{\Psi_{-n_{\rm v}\nu}}\int d^2x e\hat\psi^\dagger {\hat\psi}
\ket{\Psi_{-n_{\rm v}\nu}}-\bra{\Psi_0}\int d^2x
e\hat\psi^\dagger {\hat\psi}\ket{\Psi_0}
=e\sum_{l=0}^{n_{\rm v}-1}(-\nu)=-n_{\rm v}\nu e.
\eqno\eq
$$
In the large $B$ limit, the single-particle state are restricted in
the lowest Landau level and the eigenstate becomes $u_{l,0}^{(0)}(x)$
of Eq.(3.6). Thus the correlation function for a quasihole state
becomes
$$
\eqalign{
\pi R_0^2
\bra{\Psi_{-n_{\rm v}\nu}}\hat\psi^\dagger(y) {\hat\psi}(x)
&\ket{\Psi_{-n_{\rm v}\nu}}
\rightarrow
\nu\sum_{l=n_{\rm v}}^{\infty}{(z_yz^*_x)^l\over l!}e^{-(
z^*_x z_x+z^*_y z_y)/2}\cr
&=\nu(1-\sum_{l=0}^{n_{\rm v}-1}{(z_yz^*_x)^l\over l!}e^{
-z_y z^*_x})\exp[-\vert z_x-z_y\vert^2+i\int
\nolimits_x\nolimits^y eA_i d\xi^i],
}
\eqno\eq
$$
as $B\rightarrow\infty$. From Eq.(3.10), $\rho^2(x,y)$ reads
$$
\eqalign{
\rho^2(x,y)\rightarrow
&1-2e^{-\xi_x\xi_y\cos\theta}\sum_{l=0}^{n_{\rm v}-1}
{(\xi_x\xi_y)^l\over l!}\cos(l\theta-
\xi_x\xi_y\sin\theta)\cr
&+e^{-2\xi_x\xi_y\cos\theta}\sum_{l,m=0}^{n_{\rm v}-1}{(\xi_x\xi_y)^
{l+m}\over l!m!}\cos(l-m)\theta,}
\eqno\eq
$$
as $B\rightarrow\infty$,
where $z_x=\xi_x e^{i\theta_x}$, $\theta=\theta_x-\theta_y$.
For $\vert \xi_x-\xi_y\vert \ll1$, $n(r)$ reads
$$
\eqalign{
n(r)&\rightarrow
{\xi^{2n_{\rm v}}/(n_{\rm v}-1)!\over
e^{\xi^2}-\sum\limits
_{l=0}\limits^{n_{\rm v}-1}\xi^{2l}/l!},\ {\rm as}\ B\rightarrow\infty,
\cr
n(0)&=n_{\rm v}.}
\eqno\eq
$$
The momentum and the angular momentum density operator are defined by
$$
\eqalign{
\hat P_i(x)&={1\over2}\{\hat\psi^\dagger(x)(-i\partial_i+eA_i(x))
\hat\psi(x)+[(i\partial_i+eA_i(x))\hat\psi^\dagger(x)]
\hat\psi(x)\},\cr
\hat L(x)&=\epsilon_{ij}x_i \hat P_j(x).}
\eqno\eq
$$
Using Eqs.(3.11) and (3.12), the angular momentum of
the vortex is given by
$$
\bra{\Psi_{-n_{\rm v}\nu}}\int d^2x\hat L(x)
\ket{\Psi_{-n_{\rm v}\nu}}=-U_0\int dx^2\rho(r)n(r)
\rightarrow
-n_{\rm v}\nu,
\eqno\eq
$$
as $B\rightarrow\infty$,
where we used a relation $\VEV{\psi^\dagger(x)\partder{}{x^i}
\psi(x)}=\lim\limits_{y\rightarrow x}\partder{}{x^i}\VEV{
\psi^\dagger(y)\psi(x)}$.

Similarly we construct a quasiparticle state of a charge $+n_{\rm v}e$
for $\nu=1/3$ as
$$
\ket{\Psi_{+n_{\rm v}\nu}}
={\hat a}^{\dagger}_{n_{\rm v},0}\cdots
{\hat a}^{\dagger}_{2n_{\rm v}-1,0}
\sum_{l_1,l_2\cdots l_N=0}^M F^{(0)}_{l_1,\cdots l_N}
{\hat a}^{\dagger}_{l_1+2n_{\rm v},0}{\hat a}^{\dagger}_
{l_2+2n_{\rm v},0}\cdots
{\hat a}^{\dagger}_{l_N+2n_{\rm v},0}\ket{0}.
\eqno\eq
$$
The state $\ket{\Psi_{+n_{\rm v}\nu}}$ gives $\nu_0=\cdots=\nu_{
n_{\rm v}-1}=0$ and $\nu_{n_{\rm v}}=\cdots=\nu_{2n_{\rm v}-1}=1$
and $\nu_{2n_{\rm v}}=\nu_{2n_{\rm v}+1}=\cdots=\nu=1/3$.
In fact the charge difference from the uniform density state is
$$
\bra{\Psi_{-n_{\rm v}\nu}}\int d^2x e\hat\psi^\dagger {\hat\psi}
\ket{\Psi_{-n_{\rm v}\nu}}-\bra{\Psi_0}\int d^2x
e\hat\psi^\dagger {\hat\psi}\ket{\Psi_0}
=e\{\sum_{l=0}^{n_{\rm v}-1}(-\nu)+\sum_{n_{\rm v}}^{2n_{\rm v}-1}
(1-\nu)\}=+n_{\rm v}\nu e.
\eqno\eq
$$
In Eq.(3.16), it is essential that $\nu$ is equal to $1/3$.
In the large $B$ limit the correlation function for a quasiparticle
state becomes
$$
\eqalign{
\pi R_0^2
\bra{\Psi_{+n_{\rm v}\nu}}\hat\psi^\dagger(y) {\hat\psi}(x)
&\ket{\Psi_{+n_{\rm v}\nu}}
\rightarrow
\nu\sum_{l=n_{\rm v}}^{\infty}(1-\delta_l)
{(z_yz^*_x)^l\over l!}e^{-(
z^*_x z_x+z^*_y z_y)/2}\cr
&=\nu(1-\sum_{l=0}^{2n_{\rm v}-1}\delta_l
{(z_yz^*_x)^l\over l!}e^{
-z_y z^*_x})\exp[-\vert z_x-z_y\vert^2+i\int
\nolimits_x\nolimits^y eA_i d\xi^i],}
\eqno\eq
$$
as $B\rightarrow\infty$, where
$$
\delta_l=\left\{\matrix{
1,&{\rm for\ }0\leq l\leq n_{\rm v}-1,\cr
-2,&{\rm for\ }n_{\rm v}\leq l\leq 2n_{\rm v}-1,\cr
0,&{\rm otherwise.}
}\right.
$$
{}From Eq.(3.17), $\rho^2(x,y)$ reads
$$
\eqalign{
\rho^2(x,y)\rightarrow
&1-2e^{-\xi_x\xi_y\cos\theta}\sum_{l=0}^{2n_{\rm v}-1}
\delta_l{(\xi_x\xi_y)^l\over l!}\cos(l\theta-
\xi_x\xi_y\sin\theta)\cr
&+e^{-2\xi_x\xi_y\cos\theta}\sum_{l,m=0}^{2n_{\rm v}-1}
\delta_l\delta_m{(\xi_x\xi_y)^
{l+m}\over l!m!}\cos(l-m)\theta,}
\eqno\eq
$$
as $B\rightarrow\infty$.
For $\vert \xi_x-\xi_y\vert \ll1$, $n(r)$ reads
$$
\eqalign{
n(r)&\rightarrow
{3\xi^{2n_{\rm v}}/(n_{\rm v}-1)!-
2\xi^{4n_{\rm v}}/(2n_{\rm v}-1)!\over
e^{\xi^2}-\sum\limits
_{l=0}\limits^{2n_{\rm v}-1}\delta_l
\xi^{2l}/l!},\ {\rm as}\ B\rightarrow\infty,\cr
n(0)&=n_{\rm v}.}
\eqno\eq
$$
Using Eqs.(3.18) and (3.19), the angular momentum of the
vortex is given by
$$
\bra{\Psi_{+n_{\rm v}\nu}}\int d^2x\hat L(x)
\ket{\Psi_{+n_{\rm v}\nu}}=-U_0\int dx^2\rho(r)n(r)
\rightarrow
+n_{\rm v}\nu,
\eqno\eq
$$
as $B\rightarrow\infty$.

\noindent
(B) {\it Neutral vortex}

We also construct a neutral vortex state for $\nu=1/3$ as
$$
\ket{\Psi_{\rm neu}}=
{\hat a}^{\dagger}_{2,0}
\sum_{l_1,l_2\cdots l_N=0}^M F^{(0)}_{l_1,\cdots l_N}
{\hat a}^{\dagger}_{l_3,0}{\hat a}^{\dagger}_
{l_2+3,0}\cdots
{\hat a}^{\dagger}_{l_N+3,0}\ket{0}.
\eqno\eq
$$
The state $\ket{\Psi_{\rm neu}}$ gives $\nu_0=\nu_1=0$, $\nu_2=1$,
$\nu_3=\nu_4=\cdots=\nu=1/3$.
The charge difference from the uniform density state is zero.
In the large $B$ limit,
$\rho^2(x,y)$ reads
$$
\eqalign{
\rho^2(x,y)\rightarrow
&1-2e^{-\xi_x\xi_y\cos\theta}\{\cos(\xi_x\xi_y\sin\theta)+\xi_x\xi_y
\cos(\theta-\xi_x\xi_y\sin\theta)\cr&-
(\xi_x\xi_y)^2
\cos(2\theta-\xi_x\xi_y\sin\theta)\}
+e^{-2\xi_x\xi_y\cos\theta}\{1+(\xi_x\xi_y)^2\cr&
+(\xi_x\xi_y)^4+
2\xi_x\xi_y\cos\theta-2(\xi_x\xi_y)^3\cos\theta-2(\xi_x\xi_y)^2
\cos2\theta\}.
}
\eqno\eq
$$
For $\vert \xi_x-\xi_y\vert \ll1$, $n(r)$ reads
$$
\eqalign{
n(r)&\rightarrow
{\xi^4(3-\xi^2)\over e^{\xi^2}-1-\xi^2+\xi^4},\ {\rm as}\
B\rightarrow\infty,\cr
n(0)&=2.}
\eqno\eq
$$
{}From Eqs.(3.22) and (3.23), one can see that the angular momentum of
the neutral vortex is zero.

We see from Eqs(3.9),(3.14),(3.16) and (3.20) that the charge $Q_{\rm v}$
and the angular momentum $L_{\rm v}$ of our vortex have a relation :
$$
eL_{\rm v}=Q_{\rm v},
\eqno\eq
$$
in the large $B$ limit.

\chapter{Fluctuation corrections}
In this section we study the fluctuation corrections to the mean
field solutions. We approximate the integration over $U$ in Eq.(2.4)
into a sum over the mean field solutions of Eq.(2.8) and the quadratic
fluctuations around the mean field solutions. That is
$$
\eqalign{
Z&=e^{-S_{\rm eff}(U_0^{(n)})}\int{{\cal D}\delta U\over\cal N}
\exp[-\{S_{\rm eff}(U_0^{(n)}+\delta U)-S_{\rm eff}(U_0^{(n)})\}]\cr
&=e^{-S_{\rm eff}(U_0^{(n)})}\int{{\cal D}\delta U\over\cal N}
\exp[-{1\over2}\Bigl(
{\partial^2 S_{\rm eff}\over\partial U^{2}}\Bigr)_n
\delta U\delta U
+O(\delta U^3)]\cr
&\sim\sum_n e^{-S_{\rm eff}(U_0^{(n)})}\int{{\cal D}\delta U\over\cal N}
\exp[-{1\over2}\Bigl(
{\partial^2 S_{\rm eff}\over\partial U^{2}}\Bigr)
_n\delta U\delta U],
}
\eqno\eq
$$
where $U_0^{(n)}$'s are solutions of Eq.(2.8) and
$\Bigl(
{\partial^2 S_{\rm eff}\over\partial U^{2}}\Bigr)_n$ is given by
$$
\eqalign{\Bigl(
{\partial^2 S_{\rm eff}\over\partial U^{2}}\Bigr)_n=&
{\delta^2 S_{\rm eff}\over\delta U(y',x';t')\delta U(x,y;t)}\biggr\vert
_{U=U_0^{(n)}}=\delta(x'-x)\delta(y'-y)\delta(t'-t)V(x-y)\cr&-
V(y'-x')D_n(y',x',t';x,y,t)V(x-y)\equiv
V({\bf 1}-D_n V),}
\eqno\eq
$$
$$
\eqalign{
D_n(y',x',t';x,y,t)&=\VEV{
\psi^\dagger(y';t')\psi(x';t')\psi^\dagger(x;t)\psi(y;t)}_{U_0^{(n)}}
\cr&\qquad\qquad\qquad-
\VEV{\psi^\dagger(y';t')\psi(x';t')}_{U_0^{(n)}}\VEV{
\psi^\dagger(x;t)\psi(y;t)}_{U_0^{(n)}}\cr&
=\VEV{\psi^\dagger(y';t')\psi(y;t)}
_{U_0^{(n)}}\VEV{\psi(x';t')\psi^\dagger(x;t)}_{U_0^{(n)}}.\cr}
\eqno\eq
$$
The functional Gaussian integration in Eq.(4.1) is calculated as
$$
\eqalign{
\int{{\cal D}\delta U\over\cal N}
\exp[-{1\over2}\Bigl(
{\partial^2 S_{\rm eff}\over\partial U^{2}}\Bigr)_n
\delta U\delta U]
&=[{\rm Det}({\bf 1}-D_n V)]^{-{1\over2}}\cr
&=\exp[-{1\over2}\Tr\ln({\bf 1}-D_n V)]=\exp[{1\over2}\sum_{j=1}^{
\infty}{1\over j}\Tr(D_n V)^j],}
\eqno\eq
$$
where
$$
\eqalign{
{1\over 2j}\Tr(D_n V)^j={1\over 2j}&\int dt_1 d^2x_1 d^2y_1
\cdots dt_j d^2x_j d^2y_j V(x_1-y_1)D_n(y_1,x_1,t_1;x_2,y_2,t_2)\cr&
\cdots V(x_j-y_j)D_n(y_j,x_j,t_j;x_1,y_1,t_1).}
\eqno\eq
$$
Especially the $j=1$ term is the first order of $V$ and only includes
the equal time propagator as
$$
\eqalign{
{1\over 2}\Tr(D_n V)&={1\over2}\int dtd^2xd^2y
\VEV{\psi^\dagger(y;t)\psi(y;t)}_{U_0^{(n)}}V(x-y)
\VEV{\psi(x;t)\psi^\dagger(x;t)}_{U_0^{(n)}}\cr&
=-{\beta\over2}\int d^2xd^2yU_0^{(n)}(x,x)V(x-y)U_0^{(n)}(y,y).}
\eqno\eq
$$
Thus the $j=1$ term is a direct interaction term.
Graphical representations of Eqs.(4.5) and (4.6) are given in Fig.1.
\FIG\fia{(a) Feynman diagram for Eq.(4.5), $j=5$.
\nextline
(b) Feynman diagram for Eq.(4.6), $j=1$.}
The energy eigenvalue
appears in the partition function as $e^{-\beta E_n^{\rm tot}}$ for
$\beta\rightarrow\infty$. Thus
$$
\eqalign{
E_n^{\rm tot}=&E^{\rm mean}_n+E^{\rm ex}_n+E^{\rm dir}_n+O(V^2),\cr
E^{\rm mean}_n&=\langle E_n\vert\int d^2x\hat\psi^\dagger h_0\hat\psi
\vert E_n\rangle=\sum_{l=0}^M\nu_l E_{l,0},\cr
E^{\rm ex}_n&={1\over2}\int d^2xd^2yU_0^{(n)}(x,y)
V(x-y)U_0^{(n)}(y,x),\cr
E^{\rm dir}_n&={1\over2}\int d^2xd^2y(U_0^{(n)}(x,x)-U_0)
V(x-y)(U_0^{(n)}(y,y)-U_0).
}\eqno\eq
$$
$E^{\rm ex}_n$ is an exchange energy and $E^{\rm dir}_n$ is a direct
energy. In $E^{\rm dir}_n$, we replace $U_0^{(n)}(x,x)$ by
$U_0^{(n)}(x,x)-U_0$, because the electron density $U_0^{(n)}(x,x)$
becomes a uniform density $U_0$ at infinity and $V(r)$ does not
couple the zero mode of the Fourier component.

{}From a dimensional analysis, $V^j$ term behaves as $B^{1-j/2}$ and
$j=1$ term is a leading term of the perturbation expansion
in the large $B$ limit.

\chapter{Numerical results}
We will solve the self-consistency condition Eq.(2.8) numerically
for the charged and neutral vortex states of Eqs.(3.8),(3.15)
and (3.21).
At the start we expand $F(p^2)$ in Eq.(2.12) around $p^2=eB$
which is the energy of the lowest Landau level and approximate as
$$
\eqalign{
F(p^2)&=F(eB)+F'(eB)(p^2-eB)
=F_0-{p^2\over 2m'},\cr
 F_0&=e^2 U_0 {\pi^{3/2}\over{\kappa\gamma^{1/2}}}\times0.88934,\cr
{1\over 2m'}&={e^2\over 8\kappa} U_0 ({\pi\over\gamma})^{3/2}
\times0.48861.
}
\eqno\eq
$$
In this approximation, the eigenvalue equation in Eq.(2.13) becomes
$$
\eqalign{
[{({\bf P}+e{\bf A})^2\over2m}&+{({\bf P}+e{\bf A}+{\bf n}
)^2\over2m'}
\rho(\bar x,x)+F_0(1-\rho(x))]_{\bar x\rightarrow x}
u_{l,0}(x)={\tilde E}_l u_{l,0}(x),\cr
u_{l,0}(x)&=v_{l,0}(r)e^{-il\theta}\ ;\ l={\rm \ integer},\cr
}
\eqno\eq
$$
where we add a constant $F_0$ to the energy $h_0$ for a simplicity and
put $\tilde E_l=E_{l,0}+F_0$. For the energy eigenvalue of the
uniform density state, we define $\tilde E^{(0)}_l$ as
$$
\tilde E^{(0)}_l=E_{l,0}^{(0)}+F_0={1\over2}({eB\over M}),
\eqno\eq
$$
where $M=mm'/(m+m')$. We use the analytic form of Eqs.(3.12),(3.19)
and (3.23) for $n_i(x)$ and solve Eq.(5.2) self-consistently with respect
to $\rho$. We restrict the form of $\rho(x,y)$ as
$$
\eqalign{
\rho(x,y)&=f(\sqrt{r_x r_y},\theta^2),\cr
f(r,0)&=\rho(r),}
\eqno\eq
$$
and calculate $\rho$ by a numerical self-consistent method
$^{\B)}$.

\noindent
(A) {\it Charged vortex for }$n_{\rm v}=1,2$

Using Eqs.(3.12) and (3.19), we obtain the numerical solutions of
$\rho$. See Figs.2 and 3.
The electron densities for the quasihole and quasiparticle
vanish at the vortex core.
\FIG\fib{$\rho_{-n_{\rm v}/3}$ for quasiholes of $n_{\rm v}
=1,2$ at $B=10[T]$. }
\FIG\figc{$\rho_{+n_{\rm v}/3}$ for quasiparticles of $n_{\rm v}
=1,2$ at $B=10[T]$.}
\FIG\figd{$\rho_{\rm neu}$ for a neutral vortex of $n_{\rm v}
=2$ at $B=10[T]$. }
Figures 5 and 6 show the energy eigenvalues $\tilde E_{l}$.
\FIG\fic{$\tilde E_l$ for quasiholes of $n_{\rm v}
=1,2$ at $B=10[T]$.
$\circ$ represents an eigenvalue for $n_{\rm v}=1$ and
$\bullet$ for $n_{\rm v}=2$.}
\FIG\fica{$\tilde E_l$ for quasiparticles of $n_{\rm v}
=1,2$ at $B=10[T]$.
$\circ$ represents an eigenvalue for $n_{\rm v}=1$ and
$\bullet$ for $n_{\rm v}=2$.}
\FIG\ficb{$\tilde E_{l,0}$ for a neutral vortex of $n_{\rm v}
=2$ at $B=10[T]$. }
For large $l$, $\tilde E_{l}$ becomes degenerate to $\tilde E_{l}
^{(0)}$. For $l\leq-1$,
$\tilde E_{l}$ belongs to higher energy levels and
is larger than $\tilde E_{l}$ for $l\geq0$ by
$O(eB/M)$.
The energy differences from the uniform density state,
$\Delta E_n\equiv E_n-E^{(0)}=\tilde E_n-\tilde E^{(0)}$,
are listed in Table I for $B=10[T]$.
The gap energy $\Delta$ is a half of the pair creation energy of
a quasihole and quasiparticle which are separated infinitely as
$$
\Delta_{n_{\rm v}/3}={1\over2}(\Delta E^{\rm tot}_{n_{\rm v}/3}+
\Delta E^{\rm tot}_{-n_{\rm v}/3}).
\eqno\eq
$$
Experimentally $\Delta$ is measured as an activation energy
by the temperature dependence of the diagonal resistivity as
$\rho_{xx}\propto e^{-\beta\Delta}$. We calculate $\Delta_
{n_{\rm v}/3}$(the unit is kelvin)
at $B=5,10,15,20[T]$. See Fig.8.
The gap energy $\Delta_{2/3}$ is about two times as large as
$\Delta_{1/3}$.

\noindent
(B) {\it Neutral vortex for }$n_{\rm v}=2$

Using Eq.(3.23) we obtain the numerical solution of $\rho$.
See Fig.4. Figure 7 shows the energy eigenvalue $\tilde E_l$.
The energy differences from the uniform density state
are listed in Table.I for $B=10[T]$. The gap energy $\Delta_{\rm neu}$
for a neutral vortex is equal to $\Delta E^{\rm tot}
_{\rm neu}$. See Fig.8.
\FIG\fid{The gap energy $\Delta_{n_{\rm v}/3}$ for a charged vortex pair
of $n_{\rm v}=1,2$
and $\Delta_{\rm neu}$ for a neutral vortex at $B=5,10,15,20[T]$. }
It seems that our neutral vortex(${\rm charge}=0,\ n_{\rm v}=2$)
is a bound state of two charged vortices(${\rm charge}=
-e/3+e/3,\ n_{\rm v}=1+1$).
The gap energy $\Delta_{\rm neu}$ is about a half of the pair creation
energy $2\Delta_{1/3}$.
Since the sum of energies for two charged
vortices is larger than the energy of
one neutral vortex ($\Delta E^{\rm tot}_{-1/3}
+\Delta E^{\rm tot}_{1/3}>\Delta E^{\rm tot}_{\rm neu}$),
it is possible to make a bound state.

\chapter{Summary and discussion}
In this paper we applied a mean field theory to the FQHE for
$\nu=1/3$ and obtained the charged and neutral vortex solutions
numerically. These solutions have fractional charges and fractional
angular momenta and have a relation (3.24).
We calculated the gap energies for these vortices.
The smallest one among the gap energies is $\Delta_{1/3}$ for the
pair of vortices which have charges $\pm e/3$. The $B$ dependence
of $\Delta_{1/3}$ is written as
$$
\Delta_{1/3}=C\cdot {e^2\over2\kappa}\sqrt{eB},
\eqno\eq
$$
and $C\sim0.09$. This result is consistent with other theoretical
calculations ($C\sim0.1$) based on the Laughlin's theory
$^{\rj),\rc)}$
and on the
exact diagonalization of the finite system$^{\rb),\ra)}$.
The experimental values for the activation energy are about half of
these theoretical values at $B=10\sim20[T]$.
This discrepancy is explained by the effect
of a thickness of 2D layer and effect of disorder
$^{\rg)}$.
Below $B\sim5[T]$ the experimental gap energy becomes very small and
the discrepancy still remains.
In Ref.5), the gap energy vanished at $B=5.5[T]$ without the thickness
effect and disorder effect.
However we obtained negative results in the present paper by calculating
fluctuation corrections. The gap energy did not vanish at $B=5\sim
20[T]$.

In addition to charged vortices we obtained a neutral vortex
which has a zero charge and a zero angular momentum.
The $B$ dependence of $\Delta_{\rm neu}$ is written as Eq.(6.1) and
$C_{\rm neu}\sim0.11$.
The neutral excitation in the FQHE was studied by means of the
single-mode approximation and the exact diagonalization of the finite
system$^{\rb),\ra)}$. These studies show that the excitation energy
has a magnetoroton minimum at finite wave vector
($C_{\rm roton}\sim0.15$).
We speculate that our neutral vortex may be a wave packet made out of
the collective mode around the roton minimum.
Recently the $q=0$ collective mode in the FQHE was observed by
inelastic light scattering$^{\pdpw)}$.
In Ref.9) it is speculated that the $q=0$ excitation may be a two-roton
bound state. We hope that the magnetoroton will be observed
by inelastic light scattering or other methods.

\ACK
I am grateful to Professor K.Ishikawa for valuable discussions.

\appendix
We see in \S 5 that the energy eigenvalues become non-degenerate
for small $l$. These cause a difficulty that many-electron
state $\ket{\Psi}$ is not the eigenstate of $H_{U_0}$ in Eq.(2.12).
We can overcome this difficulty by using an arbitrariness in the
limiting procedure of discretized time slices. In this appendix we
mainly use the notation of Ref.6).

The partition function can be written as
$$
Z=\Tr(e^{-\beta \hat H})=\lim_{\epsilon\rightarrow0}\Tr[1-
\hat\psi^\dagger K\hat\psi-{1\over2}\hat\psi^\dagger\hat\psi^\dagger
V\hat\psi\hat\psi]^{\bar N},
\eqno\eq
$$
where $\beta={\bar N}\epsilon$ and we use the shorthand notations :
$$
\eqalign{
\hat\psi^\dagger K\hat\psi&=\int d^2x\hat\psi^\dagger(x)
{({\bf P}+e{\bf A})^2\over2m}\hat\psi(x),\cr
\hat\psi^\dagger\hat\psi^\dagger
V\hat\psi\hat\psi&=\int d^2xd^2y\hat\psi^\dagger(x)\hat\psi^\dagger(y)
V(x-y)\hat\psi(y)\hat\psi(x).}
\eqno\eq
$$
Using an auxiliary field $\sigma$, $Z$ becomes
$$
\eqalign{
Z&=\lim_{\epsilon\rightarrow0}\int{{\cal D}\sigma\over\cal N}e^{-
{\epsilon\over 2}\sum_k\sigma_k\sigma_k}\Tr[\prod
\limits_{k=1}\limits^{\bar N}(1-\epsilon
\hat\psi^\dagger K\hat\psi+\epsilon\hat\psi^\dagger\sigma_k v
\hat\psi-{\epsilon^2\over2}\sigma_k\hat\psi^\dagger\hat\psi^\dagger
V\hat\psi\hat\psi\sigma_k)],\cr
{\cal N}&=\int {\cal D}\sigma e^{-
{\epsilon\over 2}\sum_k\sigma_k\sigma_k},}
\eqno\eq
$$
where
$$
\eqalign{
\sigma_k\sigma_k&=\int d^2xd^2y\sigma_k(x,y)\sigma_k(y,x),\cr
\hat\psi^\dagger\sigma_k v\hat\psi&=\int d^2xd^2y
\hat\psi^\dagger(x)\sigma_k
(x,y)v(x,y)\hat\psi(y),\cr
\sigma_k\hat\psi^\dagger\hat\psi^\dagger
V\hat\psi\hat\psi\sigma_k&=\int d^2xd^2yd^2x'd^2y'
\sigma_k(x',y')\hat\psi^\dagger(x')\hat\psi^\dagger(y)
V(x-y)\hat\psi(y')\hat\psi(x)\sigma_k(y,x).}
\eqno\eq
$$
The suffix $k$ specifies a discrete time slice. We can choose a function
$v(x,y)$ in Eq.(A.3)
arbitrarily because the term $\hat\psi^\dagger\sigma_k v
\hat\psi$ is linear in $\sigma$ and vanishes after integrating over
$\sigma$. For a general $v$ we cannot exchange the order of the limit
$\epsilon\rightarrow0$ and integration $\int{\cal D}\sigma$.
Therefore we have to integrate over $\sigma$ for a finite $\epsilon$
and take a limit $\epsilon\rightarrow0$ in the end.
With a fixed $\sigma$ the $\epsilon^2$-term seems not to contribute to
Eq.(A.3). However
the integration with a Gaussian factor $e^{-{\epsilon\over2}\sum
\sigma_k\sigma_k}$ makes $\sigma$ become of order $\epsilon^{-
{1\over2}}$. Thus the $\epsilon^2$-term becomes of order $\epsilon$ by
integration $\int{\cal D}\sigma$ and is not negligible.
We change the integral variable as
$$
\sigma_k(x,y)=\sigma^{(0)}(x,y)+\xi_k(x,y),
\eqno\eq
$$
where $\sigma^{(0)}$ is a time-independent mean field solution which
is determined in the following.

The mean field $\sigma^{(0)}$ is order $\epsilon^0$ and $\sigma_k$
is order $\epsilon^{-{1\over2}}$, then $\xi_k$ is also order
$\epsilon^{-{1\over2}}$.
Hence, $Z$ becomes
$$
\eqalign{
Z&=\lim_{\epsilon\rightarrow0}\int{{\cal D}\xi\over\cal N}e^{-
{\epsilon\over 2}\sum\limits_k(\sigma^{(0)}+\xi_k)
(\sigma^{(0)}+\xi_k)}\Tr[\prod\limits
_{k=1}\limits^{\bar N}(1-\epsilon
\hat\psi^\dagger K\hat\psi
\cr&\qquad\qquad\qquad\qquad+\epsilon\hat\psi^\dagger
(\sigma^{(0)}+\xi_k) v
\hat\psi-{\epsilon^2\over2}\xi_k\hat\psi^\dagger\hat\psi^\dagger
V\hat\psi\hat\psi\xi_k)],\cr
&=\lim_{\epsilon\rightarrow0}\int{{\cal D}\xi\over\cal N} e^{-
S_{\rm eff}(\sigma^{(0)},\xi)}.}
\eqno\eq
$$
$\sigma^{(0)}$ is fixed by the stationary condition :
$$
\partder{S_{\rm eff}(\sigma^{(0)},\xi)}{\xi}\biggr\vert_{\xi=0}
=0.
\eqno\eq
$$
In the limit ${\epsilon\rightarrow0}$, $\sigma^{(0)}$ is given by
$$
\sigma^{(0)}(x,y)=v(y,x)\VEV{\psi^\dagger(y)\psi(x)}_{\sigma^{(0)}},
\eqno\eq
$$
$$
\eqalign{
\VEV{\psi^\dagger(y)\psi(x)}_{\sigma^{(0)}}&\equiv
\Tr[\hat\psi^\dagger(y)\hat\psi(x)e^{-\beta\hat H_{\sigma^{(0)}}}]/
\Tr[e^{-\beta\hat H_{\sigma^{(0)}}}],\cr
\hat H_{\sigma^{(0)}}=\int d^2x\hat\psi^\dagger K\hat\psi&-
\int d^2xd^2y\hat\psi^\dagger(x)\VEV{\psi^\dagger(y)\psi(x)}_
{\sigma^{(0)}}v(x,y)v(y,x)\hat\psi(x)\cr&
+{1\over2}\int d^2xd^2y v(x,y)v(y,x)\VEV{\psi^\dagger(y)\psi(x)}_
{\sigma^{(0)}}\VEV{\psi^\dagger(x)\psi(y)}_
{\sigma^{(0)}}.}
\eqno\eq
$$
We choose $v(x,y)$ as
$$
v(x,y)v(y,x)=V(x-y)+\sum\limits_{l=0}\limits^M
\Delta E_{l,0}u^*_{l,0}(y)u_{l,0}(x)\Bigr/
\VEV{\psi^\dagger(y)\psi(x)}_{\sigma^{(0)}}.
\eqno\eq
$$
Using the ansatz form (2.10) for $\VEV{\psi^\dagger(y)\psi(x)}$,
the energy eigenvalue equation for a single-particle becomes
$$
[K-F(({\bf P}+\hbox{\bal})^2)\rho({\bar x},x)-\Delta E_{l,0}
]_{\bar x\rightarrow x}u_{l,0}(x)=
E_{l,0}^{(0)}u_{l,0}(x),
\eqno\eq
$$
where we choosed $\Delta E_{l,0}$ as
$$
\Delta E_{l,0}=E_{l,0}-E_{l,0}^{(0)}.
\eqno\eq
$$
$E_{l,0}^{(0)}$ is the eigenvalue for a uniform density state and is
independent of $l$. Thus the energy eigenvalues become degenerate
in $l$ and the state $\ket{\Psi}$ of Eq.(2.14) becomes eigenstate
of $\hat H_{\sigma^{(0)}}$ of Eq.(A.9).
Since a difference between the eigenvalue equation in Eqs.(2.13) and
(A.11) is only a constant $\Delta E_{l,0}$, the eigenfunction
$u_{l,0}(x)$ is common in Eqs.(2.13) and (A.11).
{}From Eqs.(A.10) and (A.11), the mean field energy and
the exchange energy
become
$$
\eqalign{
E'^{\rm mean}_n&=\sum_{l=0}^M\nu_l E_{l,0}^{(0)},\cr
E'^{\rm ex}_n&={1\over2}\int d^2xd^2yv(x,y)v(y,x)
\VEV{\psi^\dagger(y)\psi(x)}_{\sigma^{(0)}}
\VEV{\psi^\dagger(x)\psi(y)}_{\sigma^{(0)}}\cr
&={1\over2}\int d^2xd^2yV(x-y)U_0^{(n)}(x,y)U_0^{(n)}
(y,x)+{1\over2}\sum_{l=0}
^M\nu_l\Delta E_{l,0}.}
\eqno\eq
$$
The quadratic fluctuation around $\sigma^{(0)}$ is calculated as
$$
\eqalign{
{\partial^2 S_{\rm eff}\over\partial\xi_k\partial\xi_{k'}}
\biggr\vert_{\xi=0}=&
\delta_{kk'}\delta(x'-x)\delta(y'-y)-\epsilon(1-\delta_{kk'})
v(y',x')D_n(y',x',t'_k;x,y,t_k)v(x,y)\cr&
+\epsilon\delta_{kk'}S(y',x';x,y),\cr
S(y',x';x,y)&=\VEV{\psi^\dagger(y')\psi^\dagger(x)V(x-y)
\psi(x')\psi(y)}_{\sigma^{(0)}}\cr
&\qquad+v(y',x')v(x,y)
\VEV{\psi^\dagger(y')\psi(x')}_{\sigma^{(0)}}
\VEV{\psi^\dagger(x)\psi(y)}_{\sigma^{(0)}}.}
\eqno\eq
$$
{}From Eq.(A.14), the direct energy included in the fluctuation correction
is written as
$$
\eqalign{
\beta E'^{\rm dir}_n&={1\over2}\Tr\epsilon\delta_{kk'}S\cr
&={\beta\over2}\int d^2xd^2y(U_0^{(n)}(x,x)-U_0)V(x-y)(U_0^{(n)}(y,y)-U_0)
+{\beta
\over2}\sum_{l=0}
^M\nu_l\Delta E_{l,0}.}
\eqno\eq
$$
{}From Eqs.(A.13) and (A.15), the total energy becomes
$$
E'^{\rm tot}_n=E'^{\rm mean}_n+E'^{\rm ex}_n+E'^{\rm dir}_n
+O(V^2)=E^{\rm tot}_n.
\eqno\eq
$$
Thus the results obtained in this paper are correct for the total
energy $E^{\rm tot}$.

\endpage
\refout
\endpage
\figout
\end

\documentstyle[12pt]{article}
\begin{document}
\pagestyle{empty}

Table I.
Energy differences from the uniform density
state at $B=10[T]$.
The unit is $eB/M$.

\

\begin{tabular}{c c c c c c}\hline\hline
$\rho_n$&$\rho_{-1/3}$&$\rho_{1/3}$&$\rho_{-2/3}$&$\rho_{2/3}$&
$\rho_{\rm neu}$\\ \hline
$\Delta E^{\rm mean}_n$ &0.03286&$-$0.15412&0.04899&$-$0.39914&$-$0.08970\\
$\Delta E^{\rm ex}_n$ &$-$0.06075&0.18320&$-$0.11480&0.39358&0.10189\\
$\Delta E^{\rm dir}_n$ &0.03391&0.02580&0.10696&0.09780&0.02356\\
\hline
$\Delta E^{\rm tot}_n$ &0.00602&0.05488&0.04115&0.09224&0.03575\\
\hline
\end{tabular}

\end{document}